\documentclass[conference]{IEEEtran}
\IEEEoverridecommandlockouts
\usepackage{cite}
\usepackage{amsmath,amssymb,amsfonts}
\usepackage{algorithmic}
\usepackage{graphicx}
\usepackage{textcomp}
\usepackage{xcolor}
\usepackage[nolist]{acronym}
\usepackage{verbatim}
\usepackage{subfigure}
\def\BibTeX{{\rm B\kern-.05em{\sc i\kern-.025em b}\kern-.08em
    T\kern-.1667em\lower.7ex\hbox{E}\kern-.125emX}}

\newcommand{\red}[1] {\textcolor[rgb]{1.0,0.0,0.0}{{#1}}}

\begin{acronym} 
\acro{3GPP}{Third Generation Partnership Project}
\acro{5GAA}{5G Automotive Association}
\acro{AWGN}{additive white Gaussian noise}
\acro{BEP}{beacon error probability}
\acro{BP}{beacon periodicity}
\acro{BSM}{basic safety message}
\acro{BSS}{basic service set}
\acro{ccdf}{complementary cumulative distribution function}
\acro{cdf}{cumulative distribution function}
\acro{CA}{collision avoidance}
\acro{CAM}{cooperative awareness message}
\acro{CAMs}{cooperative awareness messages}
\acro{CAV}{connected and automated vehicle}
\acro{CBR}{channel busy ratio}
\acro{C-ITS}{cooperative-intelligent transport systems}
\acro{CR}{channel occupation ratio}
\acro{CSD}{cyclic shift diversity}
\acro{CSI}{channel state information}
\acro{CSMA/CA}{carrier sensing multiple access with collision avoidance}
\acro{C-V2X}{cellular-\ac{V2X}}
\acro{D2D}{device-to-device}
\acro{DA}{data age}
\acro{DCC}{distributed congestion control}
\acro{DCF}{distributed coordination function}
\acro{DCM}{Dual carrier modulation}
\acro{DENM}{decentralized environmental notification message}
\acro{DMRS}{demodulation reference signal}
\acro{DSRC}{dedicated short range communication}
\acro{EDCA}{enhanced distributed coordination access}  
\acro{eNB}{evolved NodeB}
\acro{FCD}{floating car data}
\acro{FD}{full duplex}
\acro{FDD}{frequency division duplex}
\acro{FEC}{forward error correction}
\acro{GNSS}{global navigation satellite system}
\acro{GPS}{global positioning system}
\acro{HD}{half duplex}
\acro{IBE}{in-band emission}
\acro{IPG}{Inter-packet gap}
\acro{ITS}{Intelligent Transportation System}
\acro{i.i.d.}{independent identically distributed}
\acro{KPI}{key performance indicator}  
\acro{LDPC}{low density parity check}
\acro{LOS}{line-of-sight}
\acro{LTE}{long term evolution}  
\acro{LTE-D2D}{\ac{LTE} with \ac{D2D} communications}  
\acro{LTE-V2V}{\ac{LTE}-vehicle-to-vehicle}
\acro{LTE-V2X}{\ac{LTE}-\ac{V2X}}
\acro{LTE-D2D}{long term evolution with device to device communications}  
\acro{MAC}{medium access control}
\acro{MCS}{modulation and coding scheme}
\acro{MIMO}{multiple input multiple output}
\acro{MRD}{maximum reuse distance}
\acro{NGV}{Next Generation V2X}
\acro{NHTSA}{National Highway Traffic Safety Administration}
\acro{NLOS}{non-line-of-sight}
\acro{NR}{new radio}
\acro{NR-V2X}{\ac{NR}-\ac{V2X}}
\acro{OBU}{on board unit}
\acro{OCB}{outside of the context of a \acl{BSS}}
\acro{OFDM}{orthogonal frequency division multiplexing}
\acro{OFDMA}{orthogonal frequency division multiple access}
\acro{pdf}{probability density function}
\acro{PAPR}{peak to average power ratio}
\acro{PD}{packet delay}
\acro{PEP}{pairwise error probability} 
\acro{PER}{packet error rate}
\acro{PHY}{physical}
\acro{PL}{path loss}
\acro{ProSe}{proximity-based services}
\acro{PRR}{packet reception ratio}
\acro{QAM}{quadrature amplitude modulation}
\acro{QoS}{quality of service}
\acro{RBP}{resource block pair}
\acro{RR}{radio resource}
\acro{RSU}{road side unit} 
\acro{SC-FDMA}{single carrier frequency division multiple access}
\acro{SCI}{sidelink control information}
\acro{SHINE}{simulation platform for heterogeneous interworking networks}
\acro{SINR}{signal to noise and interference ratio}
\acro{SNR}{signal to noise ratio}
\acro{SPS}{semi-persistent scheduling}
\acro{SRS}{sounding reference signal}
\acro{STBC}{space time block codes}
\acro{SV}{smart vehicle}
\acro{TB}{transport block}
\acro{TDD}{time division duplex}
\acro{TTI}{transmission time interval}
\acro{UE}{user equipment}
\acro{URLLC}{ultra reliable and low latency communications}
\acro{UTDOA}{uplink time difference of arrival}
\acro{V2C}{vehicle-to-cellular} 
\acro{V2N}{vehicle-to-network}
\acro{V2I}{vehicle-to-infrastructure} 
\acro{V2P}{vehicle-to-pedestrian}
\acro{V2R}{vehicle-to-roadside}
\acro{V2V}{vehicle-to-vehicle} 
\acro{V2X}{vehicle-to-everything} 
\acro{WAVE}{wireless access in vehicular environment}
\end{acronym}

\begin{document}

\title{Co-channel Coexistence:\\Let ITS-G5 and Sidelink C-V2X Make Peace}

\author{
	\IEEEauthorblockN{Alessandro Bazzi}
	\IEEEauthorblockA{\textit{University of Bologna, DEI, Italy} \\
	alessandro.bazzi@unibo.it}
	\vspace*{-0.7cm}
\and
	\IEEEauthorblockN{Alberto Zanella}
	\IEEEauthorblockA{\textit{CNR-IEIIT, Italy} \\
	alberto.zanella@cnr.it}
	\vspace*{-0.7cm}
\and
	\IEEEauthorblockN{Ioannis Sarris}
	\IEEEauthorblockA{\textit{u-blox, Greece} \\ 
    ioannis.sarris@u-blox.com}
    \vspace*{-0.7cm}
\and
	\IEEEauthorblockN{Vincent Martinez}
	\IEEEauthorblockA{\textit{NXP, France} \\
	vincent.martinez@nxp.com}
\vspace*{-0.7cm}
}

\maketitle

\begin{abstract}\footnote{©2020 IEEE.  Personal use of this material is permitted.  Permission from IEEE must be obtained for all other uses, in any current or future media, including reprinting/republishing this material for advertising or promotional purposes, creating new collective works, for resale or redistribution to servers or lists, or reuse of any copyrighted component of this work in other works.\\Accepted version, to be presented at IEEE ICMIM 2020.}
In the last few years, two technologies have been developed to enable direct exchange of information between vehicles. These technologies, currently seen as alternatives, are ITS-G5, as commonly referred in Europe, and sidelink LTE-vehicle-to-everything (LTE-V2X) (one of the solutions of the so-called cellular-V2X, C-V2X). For this reason, the attention has been mostly concentrated on comparing them and remarking their strengths and weaknesses to motivate a choice. Differently, in this work we focus on a scenario where both are used in the same area and using the same frequency channels, without the assistance from any infrastructure. Our results show that under co-channel coexistence the range of ITS-G5 is severely degraded, while impact on LTE-V2X is marginal. Additionally, a mitigation method where the CAM data generation is constrained to periodical intervals is shown to reduce the impact of co-channel coexistence, with less degradation on ITS-G5 performance and even improvement for LTE-V2X.
\end{abstract}


\acresetall

\section{Introduction}

The upcoming revolution promised by \acp{CAV} will require reliable connectivity and will need direct communications in order to cope with scenarios where coverage by  infrastructure is not available or its intervention adds unacceptable delay. Two main  technologies have been developed in the last decade to this aim. One belongs to the Wi-Fi family and is based on the IEEE 802.11p protocol, published in 2010, which in Europe is used by the so-called ITS-G5. After about ten years of tests, including some large scale trials, ITS-G5 is considered a mature technology, which is ready for deployment. The other technology is the sidelink LTE-vehicle-to-everything (LTE-V2X)\acused{LTE-V2X}\acused{LTE}\acused{V2X}, as defined by 3GPP in Release 14 (2016-2017), and is often referred to as \ac{C-V2X} together with other solutions (uplink/downlink LTE-V2X, uplink/downlink 5G-V2X, upcoming sidelink 5G). More precisely, the ad-hoc mode of the latter, known as Mode~4, does not require assistance by the infrastructure and is thus addressed in this work (opposed to Mode~3, which requires the supervision of a central entity).  

One of the reasons of the relatively limited roll-out of V2X equipped vehicles as of today might be attributed to regulatory uncertainty. Car makers are seeking long-term viability before making heavy investments in a voluntary deployment model. 

The two technologies are targeting the reserved ITS bandwidth at 5.9~GHz and are mostly viewed as alternatives. Studies have been conducted to identify the use of these technologies in separate channels, especially focusing on the reciprocal interference and on potential strategies for the dynamic use of the channels (e.g. one technology might have priority in one channel and not in another, occupying the latter only if this is not already in use).

However, for both ITS-G5 and sidelink LTE-V2X Mode 4 (hereafter sometimes called LTE-V2X for simplicity) the multiple-access scheme is distributed and is based on sensing of the wireless channel. Thus, at least in principle, they could coexist in the same channel, provided that they share a minimum set of rules in their channel access schemes.
Based on this and on the fact that the ITS spectrum has limited channels, it has indeed been requested by regulators like CEPT to study the potential co-channel coexistence as an alternative to a band-split.

In this work, the possibility that the two technologies share the same channel is investigated. Results are obtained through simulations performed in a highway scenario, with focus on the transmission of status messages, a.k.a. \acp{CAM}. 
Coexistence is addressed by assuming that a variable portion of vehicles is equipped with one technology and the rest with the other; the realistic situation of non-periodic traffic is also compared with a case where limitations are imposed to the packet generation rules to improve reciprocal avoidance.

\section{Technologies in brief}

\subsubsection{ITS-G5} It is based at the PHY and MAC on IEEE 802.11p, now part of IEEE 802.11-2016.
IEEE 802.11p/ITS-G5 describe the \ac{PHY} and \ac{MAC} level protocols, which rely 
on \ac{OFDM} and \ac{CSMA/CA} respectively. ITS-G5 is an asynchronous ad-hoc protocol, by opposition to LTE-V2X which is synchronous with fixed time intervals. In the former, each time a station needs to transmit, it senses the medium for a given duration. If the channel is sensed busy, the station waits for the conclusion of the ongoing transmission and then senses the channel again at some random interval to reduce the probability that other stations start transmitting at the same time. Each transmission is performed using the full channel bandwidth and an \ac{MCS}, which is selected between eight options. In V2X, the majority of messages are sent in broadcast mode and thus an acknowledgment by the receiver(s) is not foreseen. 
Being based on \ac{CSMA/CA}, the time to access the channel (access delay) is usually very short, although theoretically unbounded. More details can be found for example in \cite{SepGozCol:J17,CamMolVinZha:J12}.

\subsubsection{Sidelink LTE-V2X Mode 4} Introduced by 3GPP in Release 14, at the \ac{PHY} layer it is based on \ac{SC-FDMA}, with a resource granularity of a \ac{TTI} in the time domain and a number of subchannels in the frequency domain. The TTI is 1~ms long and each subchannel is a multiple of 180~kHz. Each packet is transmitted using one or more subchannels, depending on the payload size and the adopted \ac{MCS}. A relevant aspect to the coexistence is that, in order to cope with the necessity to switch from transmission to reception, one OFDM symbol out of the 14 every TTI is left unused ($\sim$71~$\mu$s every 1~ms), thus the channel appears to be intermittently free from signals even when continuously used by LTE-V2X nodes.

The allocation process performed at the MAC layer has been designed assuming periodic messages and relies on  an \ac{SPS} approach based on a sensing mechanism performed at PHY. With a time window of 1 second, each station measures the received power in each subchannel and probes the received control messages (that are sent in parallel with the data messages). Based on this information, at given instants the resource to be adopted is randomly chosen between the 20\% expected to be less utilized and then kept for some duration (hence called SPS). This period is randomly chosen following an articulated algorithm and a parameter, which is set by the network to be within 0 and 0.8 (called keep probability). By increasing this value, the allocation can be more stable, thus reducing inaccuracy in the sensing process, but might also cause that wrong allocations, causing collisions, which may last for intervals that are even in the order of seconds. 
Insights can be found for example 
in \cite{BazCecZanMas:J18,MolGozSep:C18,TogSaiMahMugFalRaoDas:C18}.

LTE-V2X has more flexibility on the coding rate and can spread a message over a longer time duration than ITS-G5, making it more robust, although at the expense of lower effective transmit rate. In addition, the organization of the radio resources in orthogonal slots imposes an upper bounded access delay. It is also worth noting that the overall performance of LTE-V2X is strongly related to the effectiveness of the allocation process \cite{ManMarHar:C19,BazZanCecMas:J19}.

\section{Simulation settings}

Results are provided through simulations performed using the open-source LTEV2Vsim simulator \cite{CecbazMasZan:C17}, modified to cope with co-coexistence. The settings detailed hereafter and summarized in Table~\ref{Tab:Settings} are adopted.

\begin{table}
\caption{Main simulation parameters and settings.
\vspace{-2mm}
\label{Tab:Settings}}
\footnotesize
\centering
\begin{tabular}{p{3.3cm}p{4.7cm}}
\hline \hline
\textbf{\textit{Common settings}} & \\
Scenario & Highway, 3+3 lanes \\
Density & 61.5 vehicles/km \\
Speed & 140 km/h \\
Beacon periodicity & 10 Hz \\
Beacon size & 350~B \\
Channels & ITS bands at 5.9 GHz \\
Bandwidth & 10 MHz \\
Transmission power & 23~dBm \\
Antenna gain (tx and rx)  & 3 dB \\
Noise figure & 6~dB \\
Propagation model & WINNER+, Scenario B1 \\
Shadowing & Variance 3 dB, decorr. dist. 25~m \\ \hline
\textbf{\textit{ITS-G5}} & \\
MCS & 2 (QPSK, 1/2), PER=0.1@SINR=3.1dB \\
Duration of the initial space & 110\;$\mu$s \cite{ETSI_TS_102_636} \\
Random backoff & $[0\div 15] \cdot 13$\;$\mu$s \cite{ETSI_TS_102_636} \\ 
Carrier sensing threshold & -65 dBm \\
\hline
\textbf{\textit{Sidelink LTE-V2X Mode 4}} & \\
MCS & 4 (QPSK, 0.33), PER=0.1@SINR=0.1dB \\
Subchannel size & 10 resource block pairs \cite{ETSI_TS_103_613} \\
Configuration & Adjacent \\
Keep probability & 0.5 \\
Suchannel sensing threshold & -110~dBm \\
\hline \hline
\end{tabular}
\end{table}

\subsubsection{Scenario and application}\label{subsec:scenario}

The highway high speed, medium load, scenario in \cite{3GPP_TR_36_885} is assumed, with 62.5 vehicles per km on average, distributed over 6 lanes (3 in each direction). The simulated road is 2~km long. The initial position and direction of each vehicle is uniformly randomly distributed and wrap-around is applied when a vehicle exits the scenario. The results presented correspond to the average values from 20 different runs.

The cooperative awareness service is assumed. Each node generates the first packet randomly with a uniform distribution between 0 and 100 ms and then on average every 100 ms. More details on packet generation are provided in Section~\ref{Sec:coexistence}. Packets contain a payload of 350 bytes, which is a value derived from \cite{Car2Car_CAMstats}.

\begin{figure} [t]
	\centering
	\includegraphics[width=0.80\linewidth,draft=false]{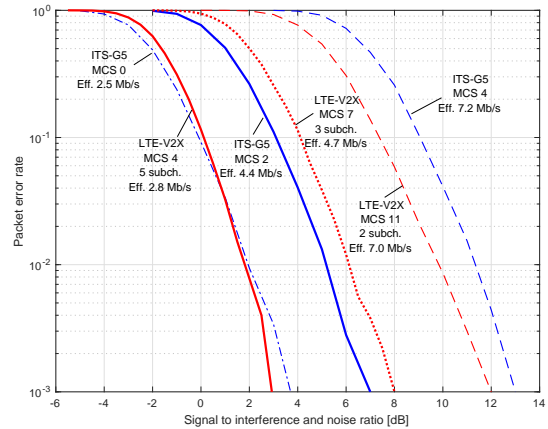}
	\caption{Packet error rate vs. average signal to interference and noise ratio in selected MCSs of ITS-G5 and LTE-V2X. Packets of 350 bytes.}
	\label{fig:PERcurves} \vskip -0.3cm
\end{figure}

\begin{figure*} [t]
	\centering
	\subfigure[ITS-G5.]{
		\includegraphics[width=0.40\linewidth,draft=false]{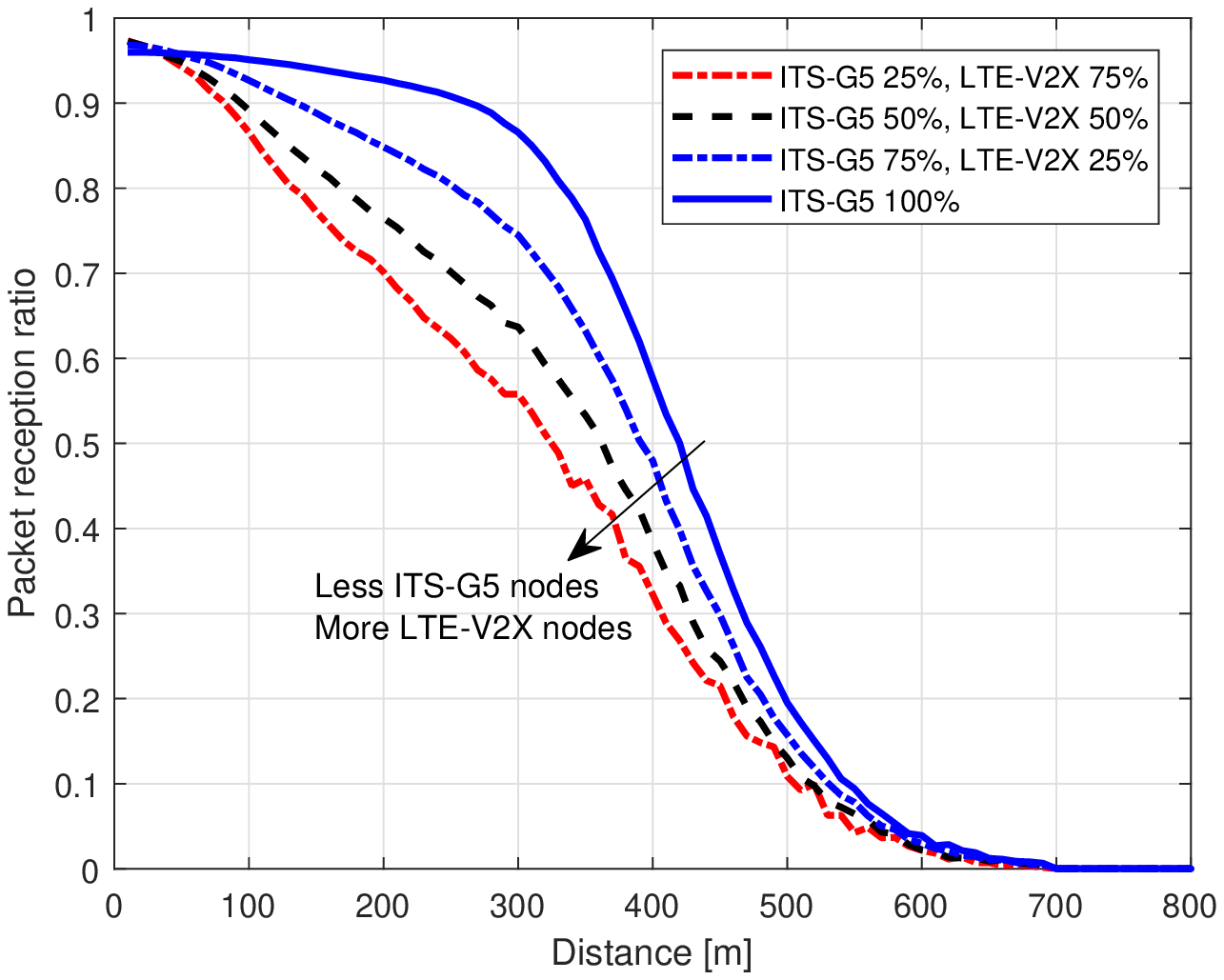}\label{fig:coexistencea}}~~~~~~~~
	\subfigure[LTE-V2X.]{
		\includegraphics[width=0.40\linewidth,draft=false]{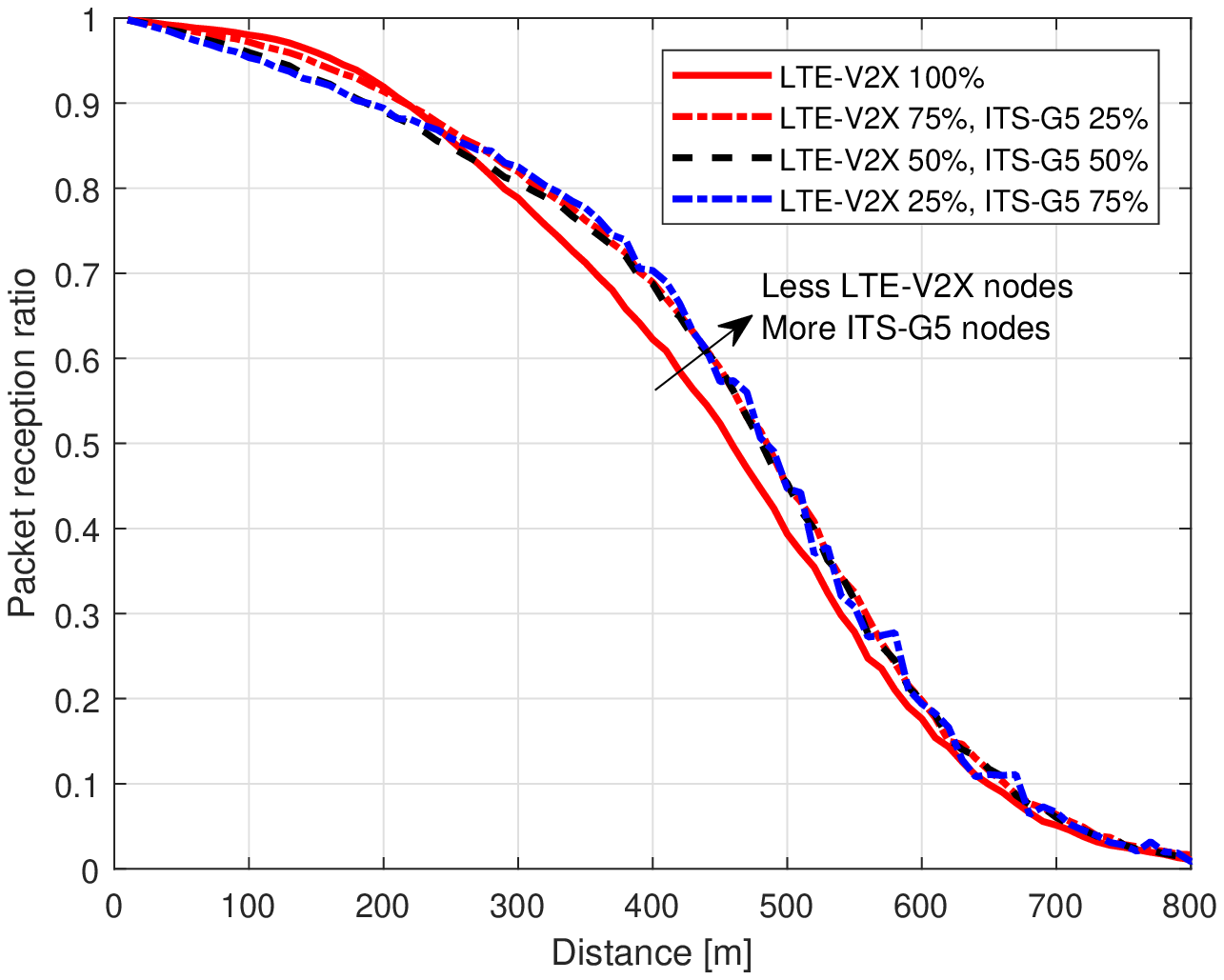}\label{fig:coexistenceb}}
	\caption{Coexistence, standard case. Packet reception ratio vs. distance with different portions of vehicles being equipped with ITS-G5 or LTE-V2X.}
	\label{fig:coexistenceStd}
\end{figure*}	

\subsubsection{Common physical layer settings}

All transmissions are performed in one or two of the channels in the 5.9 GHz ITS bands, each being 10 MHz wide. The specific channels that are used seem irrelevant to the scope of the paper.
Transmission power of 23~dBm,\footnote{In this work, LTE-V2X transmissions use all subchannels. Please note that if only a portion of the subchannels was to be used, the power should have been reduced accordingly to maintain the same power density.} 3~dB antenna gain\footnote{An antenna gain of 3~dB appears as a realistic average value over the Azimuth plane. An example antenna is the MobileMark SMW314
.} at both transmission and reception and 6~dB noise figure\footnote{A noise figure of 6 dB is for example indicated for the NXP RoadLINK SAF5400 modem
.} are considered. 
Line-of-sight (LOS)\acused{LOS} conditions are assumed. As suggested in \cite{3GPP_TR_36_885}, the path loss is modeled through the WINNER+, scenario B1, with correlated log-normally distributed shadowing, characterized by a standard deviation of 3~dB and a decorrelation distance of 25~m. 
For each potentially received packet, the average \ac{SINR} is calculated by taking into account the interference from all the other nodes. Once the SINR is calculated, the correct reception of each packet is statistically drawn from appropriate \ac{PER} vs. SINR curves, which take into account the impact of small-scale fading. In this work, the curves presented in \cite{ERMTG3720036003} and reported in Fig.~\ref{fig:PERcurves} are used, obtained assuming 2-antenna diversity at the receiver and the other settings specified in Table~\ref{Tab:Settings}. 

\subsubsection{Output metric} Results are obtained in terms of \textit{\ac{PRR}}, which is computed as the average ratio between the correctly decoded \ac{CAM} and the overall number of vehicles at a given distance.

\section{Results with the standard protocols}\label{Sec:coexistence}

ETSI specifies in \cite{3GPP_EN_302_637_2} a list of rules that trigger the generation of a new CAM, such as a change in position by more than 4 meters, a change in headings or a change in instantaneous speed. Therefore, the time intervals between packets are rarely constant and not exactly the same for all vehicles. For example, even if traveling approximately at 140 km/h, in reality some will be driving slower, some faster. In order to capture these small but relevant variations in the simulations, packets are periodically transmitted by each ITS-G5 station, assuming a periodicity which is randomly chosen between 0.095 and 0.105~s. A fixed periodicity of 0.1~s is adopted for LTE-V2X nodes, as the resources organized in TTIs restrict packet generation to specific intervals. This setting is a simplification due to the synchronized organization of LTE resources and further studies are indeed required in order to understand the implication of a correct CAM generation.

Results under these settings are shown in Figs.~\ref{fig:coexistencea} and~\ref{fig:coexistenceb}, which report the \ac{PRR} of ITS-G5 and LTE-V2X, respectively, varying the transmitter-receiver distance. In all figures, curves related to different distributions of vehicles over the two technologies are shown. 

As observable, the impact of LTE-V2X on ITS-G5 communications (Fig.~\ref{fig:coexistencea}) is remarkable. For example, the maximum distance of ITS-G5 to achieve PER above 0.9 falls from more than 250~m if all vehicles are equipped with ITS-G5 to nearly 100~m when 50\% of them adopt LTE-V2X. 

Differently, from the LTE-V2X point-of-view  (Fig.~\ref{fig:coexistenceb}) performance remains similar irrespective of the distribution among the two technologies. On the one hand, the ITS-G5 stations perform sensing before transmission and thus avoid interfering in the case of active LTE-V2X stations within range; on the other hand, the presence of ITS-G5 devices tend to make LTE-V2X selecting the preferable 20\% of resources within a smaller pool and thus increase the probability of selecting the same resource and collide.
A higher proportion of ITS-G5 stations results in a small improvement at large distances, due to the former effect, and a slight performance reduction at small distances, due to the latter.

\begin{figure*} [t]
	\centering
	\subfigure[ITS-G5.]{
		\includegraphics[width=0.40\linewidth,draft=false]{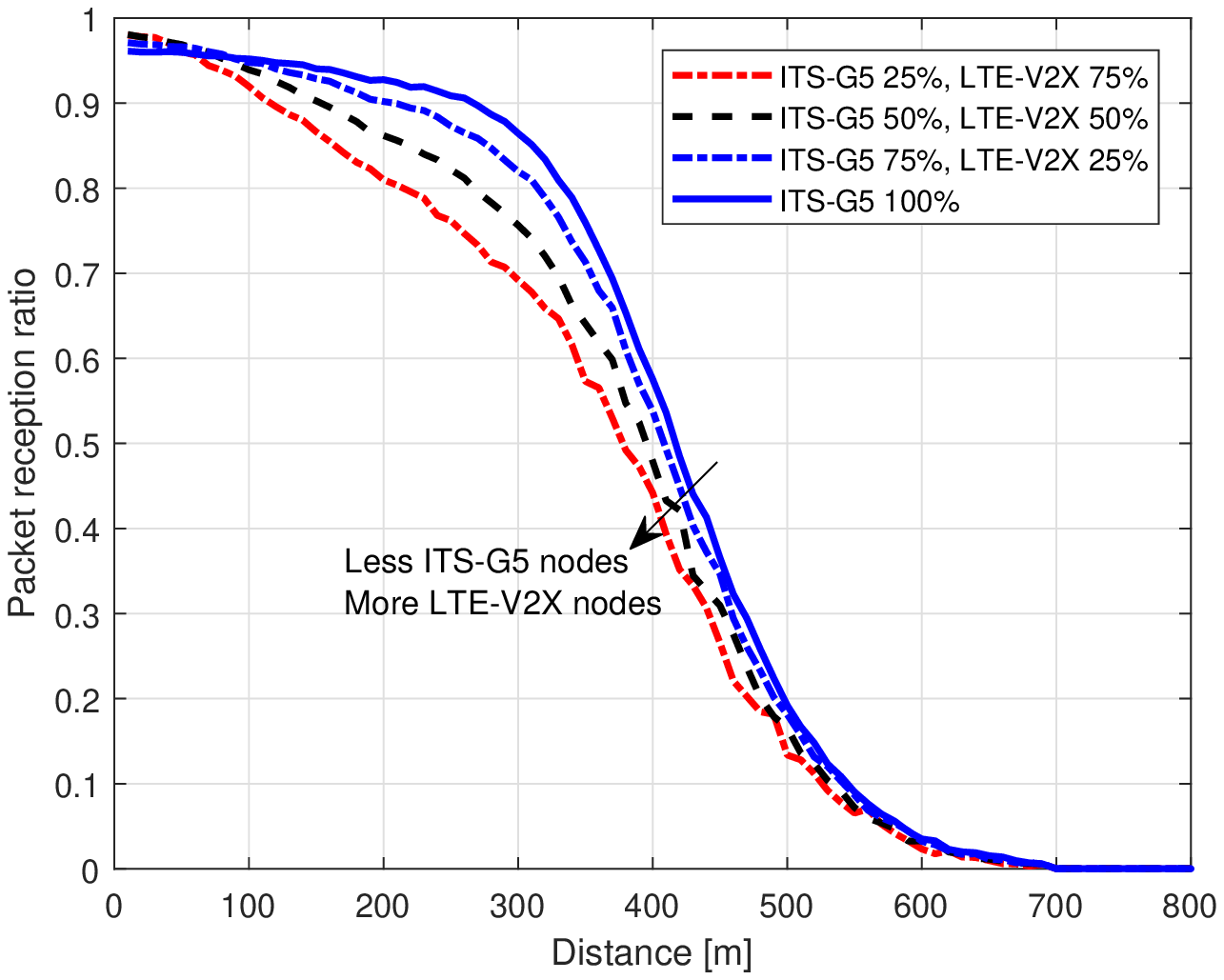}\label{fig:coexistencec}}~~~~~~~~
	\subfigure[LTE-V2X.]{
		\includegraphics[width=0.40\linewidth,draft=false]{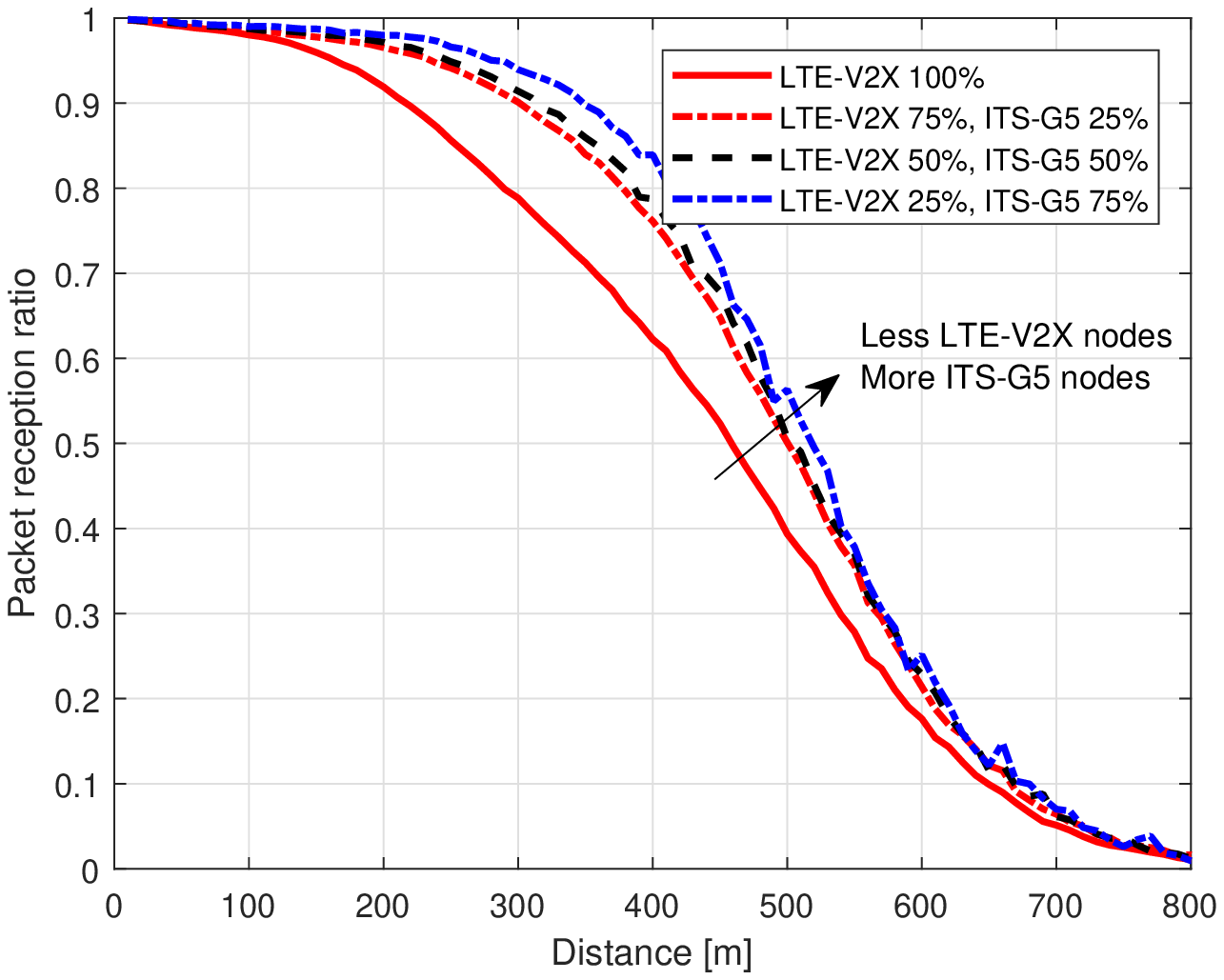}\label{fig:coexistenced}}
	\caption{Coexistence, constrained periodicity. Packet reception ratio vs. distance with different portions of vehicles being equipped with ITS-G5 or LTE-V2X.}
	\label{fig:coexistenceMod}\vskip -0.3cm
\end{figure*}

\section{Results constraining the packet generation}

The main reason for the strong impact of LTE-V2X interference on ITS-G5 is that LTE-V2X is unable to correctly estimate the use of the channel by ITS-G5 stations. In Fig.~\ref{fig:coexistenceMod}, the performance is shown, assuming a non-standard constraint to packet generation.

More specifically, it is assumed that the packets, instead of respecting the standard in \cite{3GPP_EN_302_637_2}, are generated by all nodes periodically, adopting a periodicity that complies with the LTE-V2X allocation mechanism.\footnote{The assumption is to convert the continuous function in \cite[Fig.~2-2]{Car2Car_CAMstats} to a step function complying with LTE-V2X frame structure. For example, if a car is travelling at 70/80~km/h, 200~ms might be used instead of 206/180~ms.} Specifically, in the simulations shown all nodes generate a new packet every 100~ms.\footnote{Please note that the generation of the first packet of each node is randomly chosen within 0 and 100~ms, thus synchronization among vehicles is not assumed, except for TTI synchronization in LTE-V2X. Also note that a constant and precise periodicity is indeed realistic, exploiting for example the timing provided by satellite positioning systems.}

Under these assumptions and given that the access delay in ITS-G5 is very limited in most cases, the LTE-V2X stations can individuate which TTIs have been mostly interfered by ITS-G5 stations in the past to predict which TTIs will be mostly interfered in the near future, thus reducing the reciprocal interference.

As can be observed in Fig.~\ref{fig:coexistencec}, where the PRR of ITS-G5 is shown varying the distance for various technology distributions, the negative effect of LTE-V2X signals on ITS-G5 is indeed reduced.

The effect of this modification appears to be even more favorable to the LTE-V2X communications, as shown in Fig.~\ref{fig:coexistenced}, where the corresponding PRR vs. distance curves of LTE-V2X are shown. With the constrained periodicity, the number of TTIs excluded by LTE-V2X is reduced with respect to the standard case, thus reducing the negative effect discussed in the previous subsection; in addition, the ability of ITS-G5 nodes to avoid interfering is enhanced by an effective avoidance also by LTE-V2X nodes, thus further improving the positive effect discussed in the previous subsection.

\section{Conclusion}

In this work, the co-channel coexistence of ITS-G5 and sidelink \ac{LTE-V2X} Mode~4 is investigated.
Assuming vehicles equipped with one or the other technology, in variable proportions, all using the same channel, results show that the range of ITS-G5 is severely degraded (approximately a factor of 2 in our scenario), while impact on LTE-V2X is marginal. In addition, constraining the data generation time proves to be an efficient mitigation technique.

Future work is planned to investigate other methods for improving the performance in the presence of co-channel coexistence and also considering other classes of packets. In particular, it will be interesting to focus on \acp{DENM}, which might be more critical as they are associated in ITS-G5 with shorter initial intervals than the gap foreseen at the end of TTI in LTE-V2X. Even though they are less frequently used, DENM are involved in life-threatening situations and are thus of crucial importance. 

\bibliographystyle{IEEEtran}
\bibliography{biblioSelf,biblioOthers,biblioStandards}

\begin{thebibliography}{10}
\providecommand{\url}[1]{#1}
\csname url@samestyle\endcsname
\providecommand{\newblock}{\relax}
\providecommand{\bibinfo}[2]{#2}
\providecommand{\BIBentrySTDinterwordspacing}{\spaceskip=0pt\relax}
\providecommand{\BIBentryALTinterwordstretchfactor}{4}
\providecommand{\BIBentryALTinterwordspacing}{\spaceskip=\fontdimen2\font plus
\BIBentryALTinterwordstretchfactor\fontdimen3\font minus
  \fontdimen4\font\relax}
\providecommand{\BIBforeignlanguage}[2]{{%
\expandafter\ifx\csname l@#1\endcsname\relax
\typeout{** WARNING: IEEEtran.bst: No hyphenation pattern has been}%
\typeout{** loaded for the language `#1'. Using the pattern for}%
\typeout{** the default language instead.}%
\else
\language=\csname l@#1\endcsname
\fi
#2}}
\providecommand{\BIBdecl}{\relax}
\BIBdecl

\bibitem{SepGozCol:J17}
M.~{Sepulcre}, J.~{Gozalvez}, and B.~{Coll-Perales}, ``Why 6 {Mbps} is not
  (always) the optimum data rate for beaconing in vehicular networks,''
  \emph{IEEE Transactions on Mobile Computing}, vol.~16, no.~12, pp.
  3568--3579, Dec 2017.

\bibitem{CamMolVinZha:J12}
C.~{Campolo}, A.~{Molinaro}, A.~{Vinel}, and Y.~{Zhang}, ``Modeling prioritized
  broadcasting in multichannel vehicular networks,'' \emph{IEEE Transactions on
  Vehicular Technology}, vol.~61, no.~2, pp. 687--701, Feb 2012.

\bibitem{BazCecZanMas:J18}
A.~Bazzi, G.~Cecchini, A.~Zanella, and B.~M. Masini, ``Study of the impact of
  {PHY} and {MAC} parameters in {3GPP C-V2V} mode 4,'' \emph{IEEE Access}, pp.
  1--1, 2018.

\bibitem{MolGozSep:C18}
R.~Molina-Masegosa, J.~Gozalvez, and M.~Sepulcre, ``Configuration of the
  {C}-{V2X} {M}ode 4 sidelink {PC}5 interface for vehicular communications,''
  in \emph{14th Conference on Mobile Ad-hoc and Sensor Networks (MSN 2018)},
  2018.

\bibitem{TogSaiMahMugFalRaoDas:C18}
B.~Toghi, M.~Saifuddin, H.~N. Mahjoub, M.~Mughal, Y.~P. Fallah, J.~Rao, and
  S.~Das, ``Multiple access in cellular {V2X}: Performance analysis in highly
  congested vehicular networks,'' in \emph{2018 IEEE Vehicular Networking
  Conference (VNC)}.\hskip 1em plus 0.5em minus 0.4em\relax IEEE, 2018, pp.
  1--8.

\bibitem{ManMarHar:C19}
A.~Mansouri, V.~Martinez, and J.~H{\"a}rri, ``A first investigation of
  congestion control for {LTE-V2X Mode 4},'' in \emph{IEEE Wireless On-demand
  Network systems and Services Conference (WONS)}, January 2019.

\bibitem{BazZanCecMas:J19}
A.~{Bazzi}, A.~{Zanella}, G.~{Cecchini}, and B.~M. {Masini}, ``Analytical
  investigation of two benchmark resource allocation algorithms for
  {LTE-V2V},'' \emph{IEEE Transactions on Vehicular Technology}, vol.~68,
  no.~6, pp. 5904--5916, June 2019.

\bibitem{CecbazMasZan:C17}
G.~Cecchini, A.~Bazzi, B.~M. Masini, and A.~Zanella, ``{LTEV2Vsim}: An
  {LTE-V2V} simulator for the investigation of resource allocation for
  cooperative awareness,'' in \emph{5th IEEE International Conference on Models
  and Technologies for Intelligent Transportation Systems (MT-ITS)}, June 2017,
  pp. 80--85.

\bibitem{ETSI_TS_102_636}
``Intelligent transport systems ({ITS}); vehicular communications;
  {GeoNetworking}; part 4: Geographical addressing and forwarding for
  point-to-point and point-to-multipoint communications; sub-part 2:
  Media-dependent functionalities for {ITS-G5},'' \emph{ETSI TS 102 636-4-2},
  2013.

\bibitem{ETSI_TS_103_613}
``Intelligent transport systems ({ITS}); access layer specification for
  intelligent transport systems using lte vehicle to everything communication
  in the 5,9 ghz frequency band,'' \emph{ETSI TS 103 613 V1.1.1}, 2018.

\bibitem{3GPP_TR_36_885}
``Technical specification group radio access network; study on {LTE}-based
  {V2X} services,'' \emph{3GPP {TR} 36.885 V14.0.0}, July 2016.

\bibitem{Car2Car_CAMstats}
``Survey on {ITS-G5 CAM} statistics,'' \emph{CAR 2 CAR Communication
  Consortium, TR2052}, 2018.

\bibitem{ERMTG3720036003}
``{PHY performance reference for TR103667 and TR103766 simulations},''
  \emph{ETSI ERMTG37(20)036003}, January 2020.

\bibitem{3GPP_EN_302_637_2}
``Intelligent transport systems {(ITS)}; vehicular communications; basic set of
  applications; part 2: Specification of cooperative awareness basic service,''
  \emph{3GPP {EN} 302.637-2 V1.3.1}, September 2014.

\end{thebibliography}

\end{document}